\documentclass{aa}
\usepackage{caption}
\usepackage{natbib}
\usepackage{amssymb}
\usepackage{enumerate}
\usepackage{graphicx}
\usepackage{rotating}
\usepackage{lscape}
\usepackage{txfonts}
\usepackage{epstopdf}
\bibpunct{(}{)}{;}{a}{}{,}
\begin{document}

\title{Identification of red supergiants in nearby galaxies with mid-IR photometry.
\thanks{Based on observations made with the Gran Telescopio
Canarias (GTC), installed in the Spanish Observatorio de El Roque de Los
Muchachos of the Instituto de Astrof\'{\i}sica de Canarias, in the
island of La Palma, and the 2.5 meter du Pont telescope in operation at Las Campanas Observatory, Chile.}}

\author{N. E. Britavskiy \inst{1,2} \and A. Z. Bonanos \inst{1} \and  A. Mehner \inst{3} \and D. Garc\'{\i}a-\'Alvarez\inst{4,5,6} \and J. L. Prieto \inst{7}
\and N. I. Morrell \inst{8}}

\offprints{N. Britavskiy}

\institute{IAASARS, National Observatory of Athens, GR-15236 Penteli, Greece\\
 \email{britavskiy@astro.noa.gr, bonanos@astro.noa.gr}
 \and
Section of Astrophysics, Astronomy \& Mechanics, Department of Physics, University of Athens, 15783, Athens, Greece
\and
ESO -- European Organisation for Astronomical Research in the Southern Hemisphere, Alonso de Cordova 3107, Vitacura, Santiago de Chile, Chile
\and
Instituto de Astrof\'{\i}sica de Canarias, Avenida V\'{\i}a L\'actea,
38205 La Laguna, Tenerife, Spain
\and
Departamento de Astrof\'{\i}sica, Universidad de La Laguna,
38205 La Laguna, Tenerife, Spain
\and
Grantecan S.\,A., Centro de Astrof\'{\i}sica de La Palma, Cuesta de San
Jos\'e,
38712 Bre\~na Baja, La Palma, Spain
\and
Department of Astrophysical Science, Princeton University, 4 Ivy Lane, Peyton Hall, Princeton, NJ 08544, USA
\and
Las Campanas Observatory, Carnegie Observatories, La Serena, Chile \\
}
\date{}

\authorrunning{Britavskiy et al.}
\titlerunning{Identification of RSGs in nearby galaxies}

\abstract
{The role of episodic mass loss in massive star evolution is one of the most important open questions of current stellar evolution theory. Episodic mass loss produces dust and therefore causes evolved massive stars to be very luminous in the mid-infrared and dim at optical wavelengths.}
{We aim to increase the number of investigated luminous mid-IR sources to shed light on the late stages of these objects. To achieve this we employed mid-IR selection criteria to identity dusty evolved massive stars in two nearby galaxies.}
{The method is based on mid-IR colors, using 3.6 $\mu$m and 4.5 $\mu$m photometry from archival Spitzer Space Telescope images of nearby galaxies and $J-$band photometry from 2MASS. We applied our criteria to two nearby star-forming dwarf irregular galaxies, Sextans A and IC 1613, selecting eight targets, which we followed up with spectroscopy.}
{Our spectral classification and analysis yielded the discovery of two M-type supergiants in IC 1613, three K-type supergiants and one candidate F-type giant in Sextans A, and two foreground M giants. We show that the proposed criteria provide an independent way for identifying dusty evolved massive stars, that can be extended to all nearby galaxies with available Spitzer/IRAC images at 3.6 $\mu$m and 4.5 $\mu$m.}
{}
%
\keywords{Galaxies: individual: IC 1613, Sex A -- stars: massive, late-type, red supergiant: general -- stars: fundamental parameters.}

\maketitle

\section{Introduction}

The role of mass loss from massive stars, especially episodic mass loss in evolved
massive stars, is one of the most important open questions in stellar evolution
theory. While the
upper limit to the masses of stars is thought to be $\sim$150 M$_\odot$ \citep{Figer,Oey}, or as
recently proposed, to be $\sim$300 $M$$_\odot$ \citep{Crowther10,Banerjee}, the observed masses of carbon-rich Wolf-Rayet stars, the predicted endpoints of their evolution \citep[see][]{Georgy2012}, do not exceed 20 $M$$_\odot$ \citep[see review by][]{Crowther07}. Therefore, one or more mechanisms must cause these
stars to shed
very large amounts of mass in a very short time (e.g. episodic mass loss, binary
evolution).
Classical line-driven wind theory \citep{Kudritzki00}, once thought to describe the mechanism responsible
for removing the
envelopes of massive stars, has been shown to be inadequate, both on theoretical grounds
\citep[due to wind clumping;][]{Owocki} and estimations based on spectral lines
\citep{Fullerton}, which require reductions in the mass-loss rates by factors of $10-20$.

The importance of episodic mass loss in massive stars has also come to the forefront
in the (core-collapse)
supernova community, following discoveries that emerged from recent untargeted supernovae
searches and amateur surveys. These searches have established the new classes of luminous core-collapse
supernovae
\citep[with M$_V < - 20$ mag;][]{Smith2007} and optical transients with
luminosities intermediate
between novae and supernovae \citep[e.g.][]{Prieto2008,Bond9}. In both cases, the presence
of circumstellar
material is inferred, implying a central role of episodic mass loss in the evolution
of massive
stars. The overluminous Type IIn SN 2010jl is a well-studied example of the first
class, with a
massive progenitor star of $>$30 M$_\odot$ surrounded by a dense circumstellar shell \citep{Smith2011,Zhang,Ofek}, which exploded in a low-metallicity galaxy \citep{Stoll11}.
\cite{Neill11} and \cite{Stoll11} have presented tantalizing evidence that overluminous supernovae occur in
low-metallicity host
galaxies, implying that such supernovae dominated the metal-poor early Universe.
Furthermore, there is evidence for mass-loss rates as high as $\approx$10$^{-2}$ M$_\odot$ yr$^{-1}$ in Type IIn supernovae \citep{Kiewe2012,Fox13}.
SN 2008S, a well-studied example of the class of intermediate-luminosity optical transients, was
found to have a
dust-enshrouded progenitor \citep[$8-10$ M$_\odot$;][]{Prieto2008} in pre-explosion {\it Spitzer}
images of the
host galaxy NGC 6946. Finally, the case of SN 2009ip involves a $50-80$ M$_\odot$ progenitor that
underwent episodic mass loss. The final activity included a series of eruptions
in 2009 and 2010
until its last observed explosion in August-September 2012, which resembled a Type IIn supernova \citep[e.g.][]{MSF13,PCI13,Smith2013}.

The supernovae and transients described above strongly suggest that episodic mass
loss in massive
stars is central to their evolution and has significant implications for the
enrichment of the
interstellar medium and the chemical evolution of the early Universe.
We have therefore initiated a survey that aims to provide a census of stars that have undergone episodic mass loss by studying luminous mid-IR sources in a number of nearby galaxies \citep{Khan10, Khan2011, Khan_carina}.
The ultimate goal of the survey is to investigate the role of episodic mass loss in massive
stars, by using the fact that (a) episodic mass loss in evolved massive stars
produces dust and therefore causes these sources to be very luminous in the
mid-infrared, and (b) there are abundant archival {\it Spitzer} images of nearby galaxies for which the infrared stellar population remains unexplored.
In this paper, we take advantage of mid-IR photometry from {\it Spitzer} archival data
and the “roadmap” provided by \cite{BMS09,BLK10} for interpreting luminous, massive, resolved stellar populations in nearby galaxies. 
These works studied 1268 and 3654 massive stars with known spectral types in the Large Magellanic Cloud (LMC) and Small Magellanic Cloud (SMC), respectively, and found that the brightest mid-IR sources in the LMC and SMC are luminous blue variables (LBVs), red supergiants (RSGs), and supergiants B[e] (sgB[e]), according to their intrinsic brightness and because they are surrounded by their own dust.

The RSGs are among the most numerous and luminous mid-IR populations of the massive stars family, and have some of the highest mass-loss rates observed in stars \citep[e.g. VY CMa, $2*10^{-4}$ M$_\odot$ yr$^{-1}$;][]{Danchi1994}. In the final stages of evolution, RSGs lose a significant fraction of their mass at rates $\dot{M} \sim 10^{-3}$ to $10^{-6}$ M$_\odot$ yr$^{-1}$ \citep{Jaeger1988,vanLoon2005,Mauron2011}.
For a long time RSGs were found to have lower temperatures and higher luminosities than predicted by the Geneva evolutionary models \citep{Massey03,MasseyOlsen03}; however, this disagreement was apparently solved by using the modern MARCS model atmospheres \citep{Levesque05}. The improvement was based on the new treatment of the opacity of molecular bands (e.g. TiO), which are dominant in the optical region of M- and K-type stars.
However, \cite{Davies13} reappraised the temperatures of red supergiants in the Magellanic Clouds, finding
temperatures from TiO lines to be systematically lower by several hundred Kelvin than those derived from the
continuum or integrated fluxes. These authors concluded that the
TiO bands do not provide reliable temperatures of RSGs and that their temperatures
are higher than previously thought.

The mean spectral type of an RSG in a galaxy is a function of metallicity, as first reported by \cite{Elias1985} and confirmed by recent studies \citep[e.g.][]{Levesque2006,LevesqueMassey12}. This effect is mainly due to the shift of the Hayashi limit to warmer temperatures with decreasing metallicity. In addition, at lower metallicities, the TiO bands are weaker, yielding earlier spectral types. The investigation of RSGs in the Local Group spans the following galaxies:
M33 \citep{Drout12,M98}, the Magellanic Clouds \citep{MasseyOlsen03}, M31 \citep{M98,M09}, and WLM \citep{LevesqueMassey12}.
All these works concluded that we have to increase the statistics of identified RSGs at a range of metallicities, because these stars are one of the keys for improving evolutionary models of the late stages of massive stars.
Recent works have selected candidate RSGs using optical criteria, mainly from $B-V$ and $V-R$ indices (see Section 4).
In this paper, we explore mid-IR selection criteria for RSGs and other types of dusty evolved massive stars.
The first mid-IR studies of red supergiants date back
half a century \citep{Danielson1965,Johnson1967}, but it was not until
the {\it Spitzer} era when systematic studies of the most luminous mid-IR populations
\citep{Buchanan2006,Kastner2008} and of extreme dust-enshrouded massive
stars \citep{Khan10,Khan2013,Thompson} began. The present work
differs in the mid-IR bands used, because we aim to provide selection criteria for RSGs and other obscured massive stars that can be applied to archival {\it Spitzer}
data that are available for a large number of nearby galaxies \citep[e.g. from SINGS;][]{Kennicutt2003}. The paper is
organized as follows: Section 2 presents the mid-IR selection criteria, observations
of selected targets, and data reduction. In Section 3, we discuss the spectroscopic
classification. Section 4 presents the discussion, and Section 5 the summary.

\section{Target selection and observations}

\subsection{Target selection}
We applied our mid-IR selection method to nearby galaxies with existing {\it Spitzer} photometry.
We selected two star-forming dwarf irregular galaxies, namely Sextans A of the Antlia-Sextans group \citep{vandenBergh1999}, with a radial velocity $RV$ = 323 km~s$^{-1}$  \citep{Courtois11} and distance = 1.32 Mpc \citep{Dolphin_sexa}, and the Local Group member IC 1613 with an $RV$ = --234 km~s$^{-1}$ \citep{Lu93} and distance = 0.73 Mpc \citep{Dolphin_ic}.
The metallicities of these galaxies are low, with $[O/H]\simeq$ -- 1.2 for Sextans A and $[O/H]\simeq$ -- 0.8 for IC 1613 \citep{Skillman1989}.

We identified a few dozen luminous ($M [3.6]$ $<$ $-$ 9 mag) candidate evolved massive stars in these two galaxies using archival {\it Spitzer} photometry from \cite{Boyer9}.
Our selection was based on mid-infrared colors and the exact criteria depend on the types of massive stars targeted.
All of our candidates were also searched in VizieR using a 2$\arcsec$ radius, for matches in other optical and near-IR photometric catalogs.
We also checked the proper motions of sources in VizieR to decrease the probability of selecting foreground objects.
In this paper, our primary aim is to identify RSGs, and therefore we selected sources that satisfied the following criteria:

\begin{enumerate}[i)]
\item $M [3.6]$ $<$ $-$ 9 mag
\item $J-[3.6]$ $>$ 1 and $[3.6]-[4.5]$ $<$ 0 \citep[based on][]{BMS09,BLK10}
\item $J-H$ $>$ 0.65 and $H-K$ $<$ 0.6 \citep[based on][]{RCV09}.
\end{enumerate}
We also selected other luminous sources such as candidate LBVs and sgB[e] as stars with $[3.6]-[4.5]$ $>$ 0.2 mag \citep{BMS09,BLK10}.
In this paper, we focus on sample of eight of our most luminous targets: one candidate LBV (IC 1613 1), a candidate yellow supergiant (Sex A 3), and six sources selected as RSG candidates.
The names, optical, near-IR, and mid-IR magnitudes and colors of the six selected stars in Sextans A and 2 in IC 1613, for which we present follow-up observations, are given in Table~\ref{Tab1}.

\subsection{Observations and data reduction}
The main goal of obtaining spectra of our program stars was to verify their high mass and extragalactic nature.
Our targets in Sextans A were observed with the 10.4 m Gran Telescopio Canarias (GTC) and the OSIRIS spectrograph in longslit mode during several nights in January 2013.
For our observations we used the  R1000R grism with the 1.2$\arcsec$ slit.
The resolving power provided by this combination was R $\approx$ 1100 with an average signal-to-noise ratio (S/N) $\approx$ 100 and the spectral range extended from 5100 \AA~to 10000 \AA.
Targets from the dwarf galaxy IC 1613 were observed with the 2.5 m du Pont telescope at Las Campanas Observatory, Chile.
The WFCCD with the 400 lines/mm grism and 1.6$\arcsec$ slit, gave a resolution of FWHM $\approx$ 8 \AA~at 6000 \AA~(R $\approx$ 800).
Characteristics of our observations, such as the coordinates, UT date, exposure time, and S/N for each object are presented in Table~\ref{Tab2}.
The reduction and extraction of the spectra were performed by standard IRAF\footnote{IRAF is distributed by the National Optical Astronomy Observatory, which is operated by the Association of Universities for Research in Astronomy (AURA) under cooperative agreement with the National Science Foundation.} procedures, that is dark subtraction, division by the flat fields, wavelength calibration, spectra extraction, continuum placement, and normalization.

\section{Spectral classification}

The analysis of the spectra of our targets was based on the determination of the following three characteristics:

\begin{enumerate}[i)]
 \item the radial velocity - we measured the RVs from the calcium II triplet lines ($\lambda\lambda$ 8498, 8542, 8662 \AA), which provide an important way to distinguish extragalactic objects from foreground ones.
 \item the spectral type - we expect the RSG candidates to have a K or M spectral type, which are simple to distinguish by the strength of their optical TiO bands.
 \item the luminosity class - we used the strength of the Ca II triplet, which is a good luminosity indicator for late-type stars, because it is sensitive to log g, but weakly dependent on metallicity \citep{M98}.
\end{enumerate}
To measure RVs, we cross-correlated spectral templates from the NASA Infrared Telescope Facility spectral library for cool stars \citep{RCV09} against our spectra using the IRAF task {\it fxcor}.
Before applying this method we first normalized the whole spectra using a fifth-order cubic spline, then we cut the Ca II region ($\lambda\lambda$ 8380 -- 8800 \AA) and normalized it again to a relative flux near 1.
The template spectral resolution was decreased to the value of the resolution of our spectra by convolving with a Gaussian function. The resulting radial velocities are presented in Table~\ref{Tab2}. The stars Sex A $(3, 4, 5, 6)$ and IC 1613 $(1 , 2)$ have radial velocities that confirm their extragalactic nature. Sex A $(1 , 2)$ have radial velocities near 0 and a weak Ca II triplet, thus we conclude that they are foreground giants.

\begin{figure*}
\resizebox{\hsize}{!}{\includegraphics{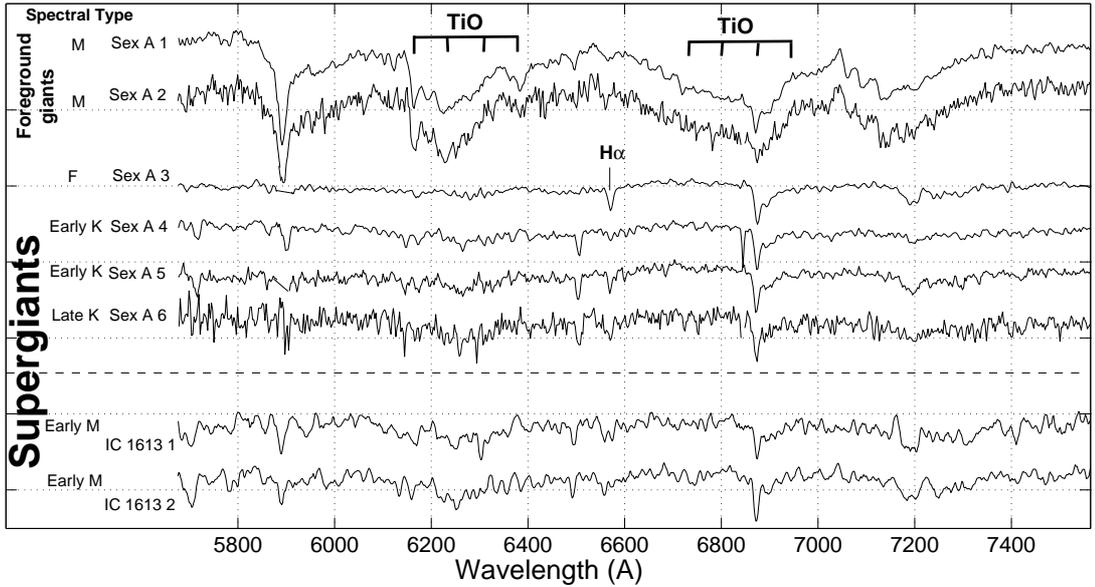}} 
\caption[]{Spectra of our eight program stars in the optical range ordered by spectral type, showing the evolution of the TiO bands from early to late spectral types.}
\label{Fig1}
\end{figure*}

We determined the spectral type with the help of the ESO UVES Paranal Observatory Project (POP)
library of high-resolution spectra across the H-R diagram \citep{Bagnulo}. This high-resolution library contains a broad range of spectral types from F2 to M6 and luminosity classes ranging from dwarfs (V) to supergiants (I). We decreased the resolution of the templates from R $\approx$ 70000 to R $=$ 1000 to match our spectra, as described above.
At this resolution it is quite difficult to distinguish the lines that are indicators of the spectral type - instead of lines, we only have blends. Nevertheless, the comparison with templates provides us with the opportunity to investigate the behavior of the TiO bands, which are very strong for K and M spectral types. The resulting spectral type classification is shown in Fig.~\ref{Fig1}.
The accuracy of our spectral type classification is at the level of distinguishing late/early spectral subtypes. Our spectra are also of sufficient quality for further luminosity classification, namely for separating the supergiants from giants.

\begin{figure*}
\resizebox{\hsize}{!}{\includegraphics{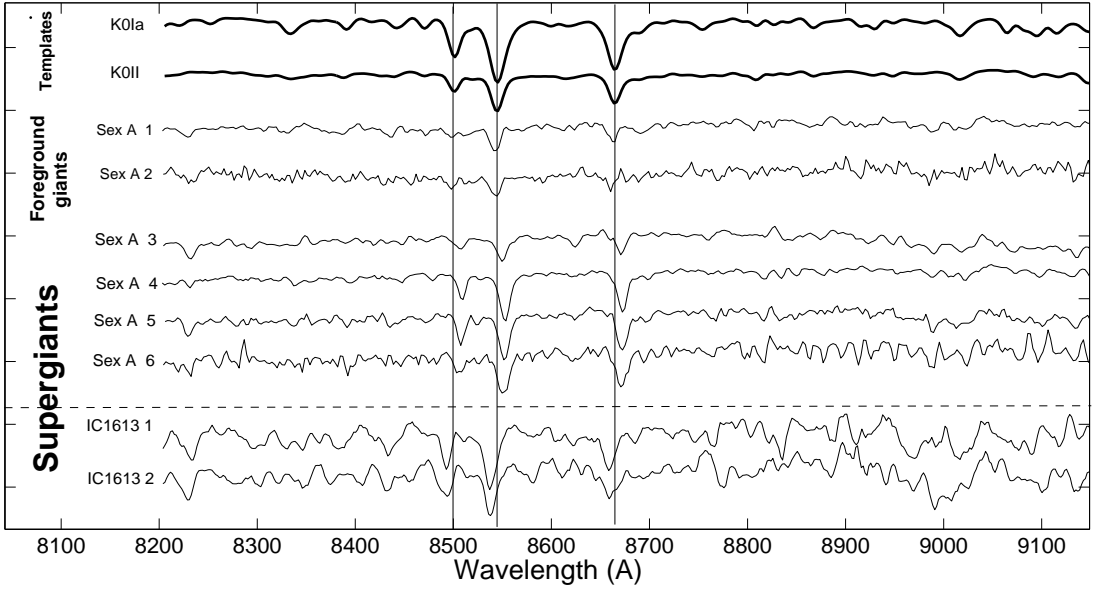}} 
\caption[]{Ca II triplet region for all the targets. The upper two objects are templates from \cite{RCV09}.
Vertical lines show the rest wavelength position of the calcium triplet ($\lambda\lambda$ 8498, 8542, 8662 \AA), relative to which the Doppler shifts of spectra are measured. We conclude that Sex A (3,4,5,6) and IC 1613 (1,2) are extragalactic objects.}
\label{Fig2}
\end{figure*}

The classification of red supergiants in the optical/near-IR region has been explored by several authors. Different criteria have been
proposed depending on the available resolution. The spectral classification of low-resolution spectra is described in \cite{Ginestet94}, \cite{Negueruela11,Negueruela12}, and \cite{M98}, who presented a classification of RSGs in M31 and M33.
However, the resolution of our spectra is not high enough to identify luminosity-sensitive features (Fe, Ti) apart from the Ca II triplet, which we used for identifying RSGs in our sample.
In Fig. \ref{Fig2} we present the Ca II triplet region for all our targets. The first two spectra (from top to bottom) are taken from the IRTF atlas \citep{RCV09}, demonstrating the difference between the strength of the Ca II features for a K supergiant and a bright giant. The next two spectra have an RV near zero therefore these objects are foreground M dwarfs or giants. The other spectra are extragalactic, as deduced from their radial velocities and depths of Ca II triplet lines. All are supergiants, apart from Sex A 3 (we discuss this object below).

\begin{figure*}
\resizebox{\hsize}{!}{\includegraphics{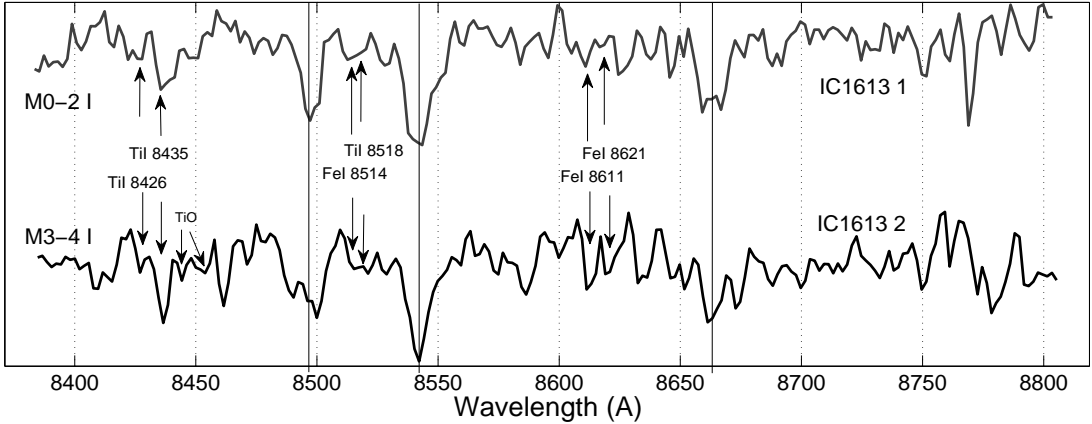}} 
\caption[]{Ca II triplet spectral region of the IC 1613 targets. Using the criteria described by \cite{NMG12}, we find a spectral type M0-2 I for IC 1613 1 and M2-4 I for IC 1613 2 (see Section 3).}
\label{Fig3}
\end{figure*}

The M-type spectra of the IC 1613 targets had a high enough quality to resolve the Fe I, Ti I, and the TiO bands and we accomplished accurate classification using the criteria described in \cite{NMG12}. The relations between the strengths of the Ti I and Fe I in the Ca II triplet region enable the determination of the spectral type.
We found several criteria for the classification of these two stars (see Fig. \ref{Fig3}):
the strength of the 8514 \AA~feature compared with Ti I 8518 \AA~is an indicator of high luminosity,
while the TiO bandhead features at 8442 \AA~and 8452 \AA~visible in IC 1613 2 are a sign of M4 type.
The fact that the line Fe I 8611 \AA~is stronger than Fe I 8621 \AA, which is usually contaminated by the 8620 \AA~DIB, indicates the extragalactic nature of these stars.
In M-type spectra the relative strength of the various Ti lines indicates how late the spectral type is.
In particular, the strength of Ti I 8426 \AA~and of the TiO features at 8432 \AA~blended with the Ti I 8435 \AA~line becomes similar to the TiO 8660 \AA~/ Ca II 8662 \AA~features and can be used as an indicator of spectral type.
We therefore find a M0--2 I classification for IC 1613 1 and M2--4 I for IC 1613 2.

We applied two methods, which are based on the Ca II triplet, to confirm our classification: (i) the investigation of the continuum level around the Ca II region and (ii) a quantitative profile analysis of the Ca II triplet.
The first method uses templates of cool stars \citep{RCV09} to compare of the Ca II region continuum shape. By shape we mean the average continuum, which consists of blends where we cannot separate the individual lines of the elements. In this step we confirm our spectral type classification because the level of the continuum around the calcium triplet region is quite sensitive to temperature, mainly because of the TiO bands, as mentioned above.
The second method of our analysis was to quantitatively analyze of the Ca II triplet profiles to confirm the luminosity classification.
We used the empirical calibration of the near-IR index CaT* \cite[defined by][]{Cenarro01}, which measures the Ca II triplet strength, corrected for the contamination from Paschen lines. This index is quite universal with small limitations on the properties of the input spectra and applies to a wide range of metallicities. The sample of targets from the du Pont telescope (the targets from IC 1613) have a resolution $FWHM = 8$ \AA, while the spectral resolution of the Cenaro spectra library, from which these indices have been determined, is $FWHM =1.5$ \AA. We do not pretend to have computed the real CaT* indices, we have measured only a quantitative characteristic value of the depth of the Ca II triplet of our sample. We calculated indices and errors by using the indexf \citep{indexf} program package.  Results are shown in Table \ref{Tab2}. For foreground giants this index is half that of supergiants and agrees with the empirical fitting-function library \citep{Cenarro02} for stars that have the same spectral type and metallicity. The CaT* index measurements therefore confirm the supergiant nature of 5 stars.

Object Sex A 3 deserves particular attention. We included this luminous mid-IR source among our targets to explore its nature, even though this object only satisfies our luminosity selection criteria. The radial velocity analysis indicates an extragalactic nature of this star, while the comparison with the ESO spectral library and the presence of H$\alpha$ reveals that this object has an F spectral type. For F-type supergiants the O I $\lambda$ 7774 line is stronger than in dwarfs due to non-LTE effects \citep{Przybilla00} and the six hydrogen Paschen series in the region of the Ca II triplet ($\lambda \lambda$ P14 8598, P12 8750, P11 8863 and the three lines contaminated by calcium lines: P13 8502, P15 8545, P16 8665) should also be visible \citep{Drout12}. However, the O I line is absent in the spectrum of Sex A 3 and the Paschen lines are not distinguishable. In addition, the Ca II triplet is weaker than the template spectra.
We conclude that this object is most likely an extragalactic giant, but to give a definitive answer about the nature of this star it is necessary to obtain a higher-resolution spectrum.
\section{Discussion}
\subsection{Comparison with other selection methods}
\begin{figure*}
\resizebox{\hsize}{!}{\includegraphics{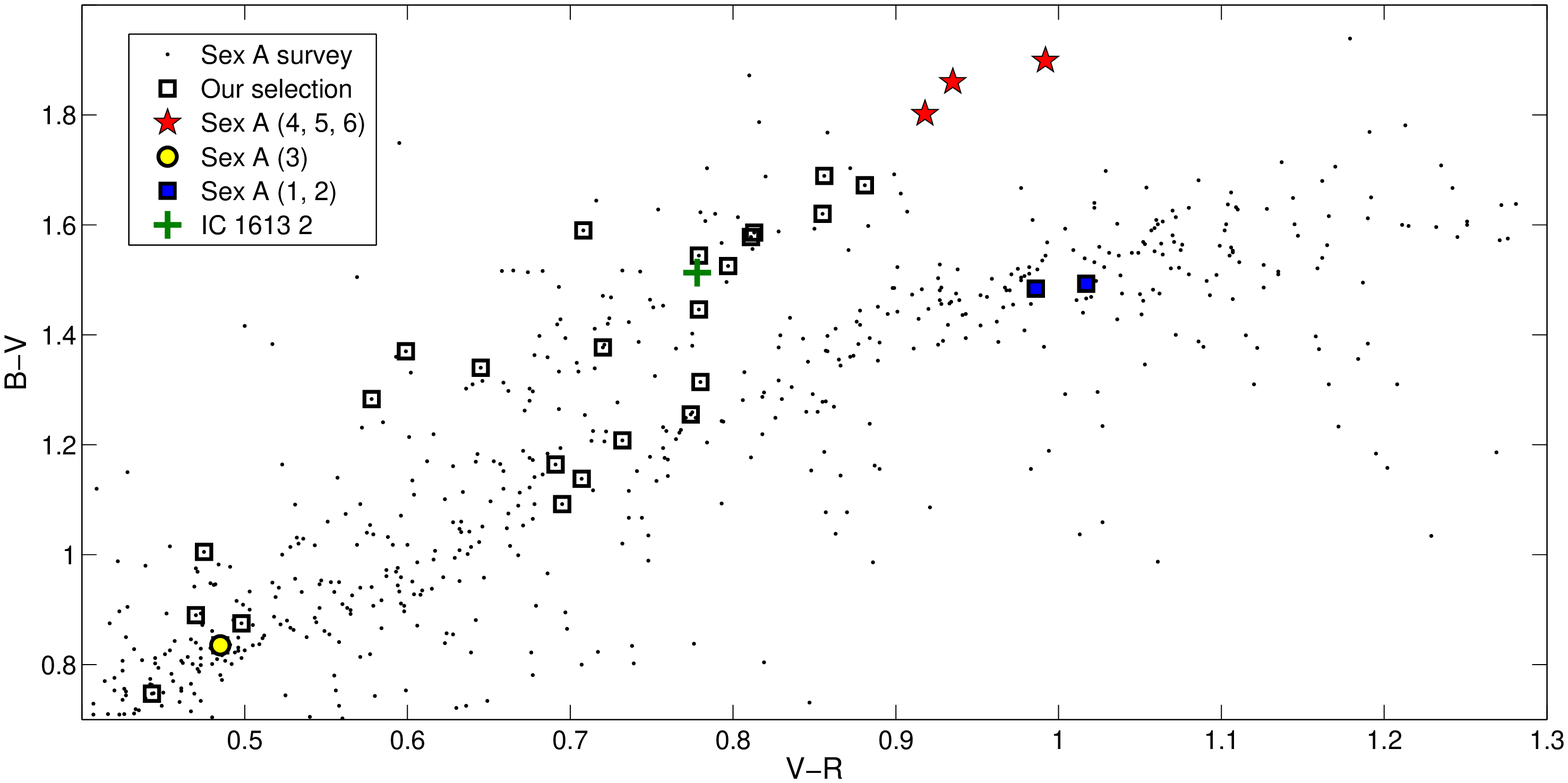}} 
\caption[h]{Two-color diagram for stars in Sex A \citep[from][]{Massey9phot},
which is a useful tool for distinguishing RSGs from foreground giants
\citep{M98}. The positions of our investigated targets are
labeled using different symbols. All targets in Sex A that satisfy our selection criteria (see Section 2.1) are marked by black squares; the majority lie on the upper, RSG branch.}
\label{Fig4}
\end{figure*}

To validate our method of target classification we used photometry from the optical $UBVRI$ survey conducted for Sextans A by \cite{Massey9phot} and unpublished photometry for IC 1613 (M. Garcia, private comm.) described in \cite{Garcia}. In Fig. \ref{Fig4} we present a $B-V$ versus $V-R$ diagram for Sextans A, indicating all our targets except for IC 1613 1, because accurate photometry does not exist for this target.
All targets that were selected as massive star candidates are marked by black squares.
\cite{M98} showed that stars with a high $B-V$ value at a given $V-R$ are expected to be supergiants, and therefore these colors are a useful tool for distinguishing supergiants from foreground giants, although in practice the method is not $100\%$ efficient.
We found that the objects Sex A 4, 5, 6, and IC 1613 2 occupy the supergiant region on this color-color diagram, while the objects Sex A 1 and 2 are found below the supergiant sequence, as expected for foreground objects.
The object Sex A 3 remains controversial. The position of this object on the diagram, compared with the location of RSGs, indicates a higher surface gravity and earlier spectral type for this object, which is consistent with an F-type giant.
Therefore we conclude that the two different approaches to selecting RSGs based on optical and mid-IR colors agree and can be used in combination with each other.
\subsection{Contamination from other luminous sources}
The most probable contaminants of normal RSGs on the mid-IR CMDs, given our
selection criteria, are the AGB stars and, in particular, the super-AGB stars \citep{Garcia-Berro, Herwig}.
\cite{Messineo12} provided new criteria to separate RSGs and LBVs from AGB stars, which are based on near-IR indices, $Q1$ and $Q2$. These indices measure the deviation from the reddening vector in the $H-K_{s}$ versus $J-K_{s}$ plane and the $J-K_{s}$ versus $K_{s}-[8.0]$ plane, respectively. We applied this technique to our sample of sources (see Table~\ref{Tab1}) and found that six targets satisfied the proposed criteria for RSGs. The range of these parameters for RSGs should be 0.1 $<$ $Q1$ $<$ 0.5 and $-1.1$ $<$ $Q2$ $<$ $1.5$.
The exceptions are the object Sex A 3, for which we have no precise [8] color, and IC 1613 1, which has values of $Q1$ $=$ $0.53$ and $Q2$ $=$ $-10.38$, typical of AGB stars ($-$1.6 $<$ $Q1$ $<$ 0.4 and $-$13 $<$ $Q2$ $<$ $-$1), but this object was selected as a candidate LBV star (the typical values are $-$0.3 $<$ $Q1$ $<$ 0.0 and $-$2.2 $<$ $Q2$ $<$ $-$0.7).
The object IC 1613 1 is a good example of the fact that different types of massive stars overlap almost on all the color magnitude diagrams (CMDs).
\cite{Mantegazza} showed that object IC 1613 1 is a blue variable with a period P $= 62.71$ days and amplitude $\Delta$V $=$ 0.35 mag. However, we found a range of values of the visual magnitude from different references, thus we expect a higher amplitude of variability for this source.
We selected this object as an LBV with $[3.6]-[4.5]$ $= 0.667$ mag \cite[from the criteria of][]{BMS09,BLK10}. \cite{Thompson} proposed that LBVs have $[3.6]-[4.5]$ $<$ 0.8 mag and AGB stars $0.5$ $<$ $[3.6]-[4.5]$ $<$ $1.5$ mag.
The general conclusion that we can make is that during the selection process it is very important to consider contamination by oxygen-rich AGB stars, because there is an overlap in color indices (especially in $[3.6]-[4.5]$) between LBVs and AGBs and in both color indices and luminosities for RSGs and super-AGB stars as well. Nevertheless, there is a spectroscopic indicator of AGB stars; the line Li I 6708 \AA~ should be visible in high-resolution spectra of O-rich AGB stars and may be used as a reference line for distinguishing AGB stars from RSGs \citep{Garc2013}.
\subsection{Variability}
Another effect that we took into account is the variable nature of massive stars \citep[e.g.][]{Kourniotis2013}.
\cite{Szczygiel} showed that the majority of the identified variable massive
stars in the LMC are RSGs, while the RSGs and AGB stars in M33 have been the subject
of variability studies in the near-IR \citep{Javadi2011} and mid-IR \citep{McQuinn2007}. RSGs and AGB stars often have a quite similar long-period variable (LPV) nature \citep{Wood1983}.
In fact, it is possible to separate RSGs from AGB stars by using the difference in variability properties, which is important for the luminosity of super-AGB stars that can contaminate our sample.
The distinction is based on the difference of absolute $K$ magnitude against variability period for these groups of stars \cite[see Fig. 2 in][]{Wood1983}.
Another difference between supergiants and AGB stars is in the amplitude of their variability.
The AGB stars have a larger variable amplitude than RSGs.
Variability does affect the color indices through a pulsation cycle, thus we may obtain different conclusions about the nature of a star depending on the observational epoch.
For instance, the object IC 1613 1 shows unusual color indices (see Table~\ref{Tab1}), probably as a result of the large variability amplitude.
The variable nature of massive stars might be connected with the phenomenon of unusual spectral type of RSGs with respect to the average spectral type among the host galaxy \citep[discussed in][]{Levesque10_phys, LevesqueMassey12}. The effect consists in a progression of the dominant spectral type toward earlier types at lower metallicities.

\begin{table*}
{\tiny
\caption[]{Photometry and color indices of the program stars.}
\label{Tab1}
\begin{tabular}{ccccccccccrrr}
\hline
Star    &$V$    &$B-V$ & $V-R$ & $M[3.6]$* &  $[3.6]-[4.5]$ & $J-[3.6]$ & $J-H$ & $H-K$    & Q1 & Q2   \\
Name    & (mag) & (mag) &(mag)  & (mag)         & (mag)  &  (mag)          &        (mag)     & (mag)  &    &     \\
\hline
Sex A 1  &  18.547 & 1.493 $\pm$ 0.008  & 1.017 $\pm$ 0.005 &$-$11.31 & 0.052 	&  1.05 $\pm$ 0.05 & 0.932 $\pm$ 0.069 & 0.175 $\pm$ 0.116	& 0.30   & 0.53       \\
Sex A 2  & 19.097  & 1.484 $\pm$ 0.011  & 0.986 $\pm$ 0.006 & $-$10.71  & 0.121 	&  1.20 $\pm$ 0.08 & 0.900 $\pm$ 0.108 & $-$0.034 $\pm$ 0.181 & 0.25   & $-$0.28   \\
Sex A 3  & 18.650  & 0.835 $\pm$ 0.006  & 0.485 $\pm$ 0.005 & $-$9.40   & 0.049 	&  0.53 $\pm$ 0.16 & 0.451 $\pm$ 0.281 & 1.632 $\pm$ 0.227	 & 0.61   & --     \\
Sex A 4    & 18.295   & 1.860 $\pm$ 0.008  & 0.935 $\pm$ 0.005 & $-$11.11  &$-$0.064	&  1.08 $\pm$ 0.05 & 0.902 $\pm$ 0.084 & 0.058 $\pm$ 0.135	 & 0.30   & 0.40      \\
Sex A 5  & 18.322   & 1.802 $\pm$ 0.008  & 0.918 $\pm$ 0.005 & $-$11.11 &$-$0.027	&  1.16 $\pm$ 0.06 & 0.763 $\pm$ 0.075 & 0.176 $\pm$ 0.114	 & 0.32   & 0.61      \\
Sex A 6  & 18.596   & 1.899 $\pm$ 0.009  & 0.992 $\pm$ 0.005 & $-$11.11 &$-$0.004	&  1.26 $\pm$ 0.06 & 0.883 $\pm$ 0.090 & 0.001 $\pm$ 0.155	 & 0.26   & $-$0.18   \\
\hline
IC 1613 1   &18.98$\dagger$ &  --          &    --                    & $-$9.52 & 0.667 	&  3.26 $\pm$ 0.05 & 0.337 $\pm$ 0.087  & 0.973 $\pm$ 0.091 & $-$0.53  & $-$10.38  \\
IC 1613 2  &17.258  & 1.513 $\pm$ 0.006 & 0.778 $\pm$ 0.011   & $-$11.02 &$-$0.009	&  0.90 $\pm$ 0.03 & 0.692 $\pm$ 0.037   & 0.212 $\pm$ 0.044  &  0.32  &  0.65   \\
\hline
\end{tabular}
\\
$\ast$ The distance modulus that we used is 25.61 $\pm$ 0.07 mag for Sextans A \citep{Dolphin_sexa}, and 24.32 $\pm$ 0.06 mag for IC 1613 \citep{Dolphin_ic}. \\
$\dagger$ Several independent estimates of magnitude and colors are available for this star (e.g. $V =$ 17.97; 19.33). Details in Section 4.2.
}
\end{table*}


\begin{table*}
{\tiny
\caption[]{Log of observations (coordinates, UT date, exposure time, signal-to-noise ratio), derived radial velocities, CaT* indices, and spectral types. RV$_{Sex~A}$ = 323 km~s$^{-1}$ and RV$_{IC~1613}$ = $-$238 km~s$^{-1}$.}
\label{Tab2}
\begin{tabular}{cccccrrccc}
\hline
Star        & R.A.   &  Dec &  UT Date     &  Exp. time & S/N & RV $\pm$ $\sigma_{RV}$ & CaT*  $\pm$ $\sigma_{CaT*}$  & Spectral & Luminosity          \\
Name     & (J2000) & (J2000)   & (HJD-2450000)&   (sec)        &    & (km~s$^{-1}$)          &                              & classification & classification\\
\hline
Sex A 1  &  10 10 58.12  & $-$04 42 08.89 & 6326.73523 & 600 & 100 &$-$10 $\pm$ 18 & 4.2 $\pm$ 1.2 & Late M   & III-V\\
Sex A 2  &  10 11 00.00  & $-$04 39 05.69 & 6320.57939 & 500 & 120 &3 $\pm$ 22 & 4.2  $\pm$ 1.1 & Late M   & III-V\\
Sex A 3  &  10 11 10.28  & $-$04 42 06.19 & 6330.52805 & 600 & 50  &273 $\pm$ 18 &  4.3  $\pm$ 1.4 & F        & III-V\\
Sex A 4  &  10 10 53.80  & $-$04 41 07.32 & 6298.62638 & 500 & 150 &368 $\pm$ 15 &  10.7 $\pm$ 2.2 & Early K  & I\\
Sex A 5  &  10 11 05.54  & $-$04 41 56.91 & 6300.71179 & 500 & 80  &352 $\pm$ 21 &  10.2 $\pm$ 2.3 & Early K  & I\\
Sex A 6  &  10 10 56.71  & $-$04 40 39.06  & 6320.54736 & 500 & 120 &305 $\pm$ 27 &  9.5  $\pm$ 2.2 & Late K   & I\\
\hline
IC 1613 1 &  01 04 40.77  & $+$02 01 10.09  & 6299.54371 & 1200& 14 &$-184$ $\pm$ 25 &  10.9  $\pm$ 2.9& M$0-2$ & I \\
IC 1613 2 &  01 04 38.52  & $+$02 00 57.60  & 6301.57033 & 1800& 24 &$-186$ $\pm$ 7 &  15.9  $\pm$ 2.5 & M$2-4$ & I \\
\hline
\end{tabular}
\\
}
\end{table*}

\section{Conclusions}
In summary, we have presented mid-IR selection criteria based on the  [3.6], [4.5] and the {\it J} band for identifying dusty evolved massive stars and particulary RSGs, which are independent and complementary to selection methods for RSGs based on optical colors.
We selected a sample of eight dusty massive star candidates from two nearby dwarf galaxies with a high star-formation rate, Sextans A and IC 1613.
We presented evidence from our spectroscopic analysis confirming that five objects from our sample are RSGs.
In addition, we compared our selection criteria with two other approaches for identifying RSGs.
We found our selection process in agreement with the optical color-color $BVR$ selection of \cite{M98} and with the near-IR color criteria of \cite{Messineo12} for separating RSGs from other types of evolved massive stars.
We caution that contamination by AGB stars does occur during the photometric selection and that RSGs and super-AGB stars are almost indistinguishable in their colors, luminosity, and optical spectra. This contamination is more prominent for Galactic objects, for which there is strong interstellar extinction and, furthermore, radial velocity measurements do not help in selecting extragalactic objects.
Combining the different selection techniques will increase the accuracy of the selection process, but we showed that our criteria are completely independent.
In future work devoted to the discovery of dusty massive stars in nearby galaxies, it would be preferable to obtain follow-up spectra in the near-IR instead of in the optical wavelength range because we only have high enough S/N at $\lambda \lambda$ 7000 $-$ 10000 $\AA$ for these red objects.
In the near-IR bands there are more features that can help in determining the luminosity class and spectral type classification.

Currently, infrared data on nearby galaxies available from {\it Spitzer} surveys \citep[e.g. SINGS;][]{Kennicutt2003} remain unexplored concerning dusty evolved massive stars. The next step is to use the {\it Spitzer} Legacy Survey of Local Group Dwarf Galaxies \citep{Boyer2011, Boyer2012}, which covers a large number of dwarf irregular galaxies and creates opportunities for identifying dusty massive stars.
Together with the other methods and approaches for identifying dust-obscured massive stars \citep{Khan10,Thompson}, our mid-IR selection criteria will allow us to carry out the census of evolved dusty massive stars that will help us understand the role of episodic mass loss phenomena in the evolution of massive stars.

\section*{Acknowledgments}
We thank the anonymous referee and D. Lennon for helpful comments that have improved the manuscript.
N. Britavskiy and A.Z. Bonanos acknowledge funding by the European Union (European Social Fund) and
National Resources under the "ARISTEIA" action of the Operational Programme "Education and Lifelong Learning" in Greece.
This research has made use of NASA's Astrophysics Data System Bibliographic Services and the VizieR catalogue access tool, CDS, Strasbourg, France.
\bibliographystyle{aa}
\bibliography{ref}

\end{document}